2009, Vol. 3, No. 3, 1124–1146

*The Annals of Applied Statistics*
2009, Vol. 3, No. 3, 1124–1146
DOI: 10.1214/09-AOAS238
© Institute of Mathematical Statistics, 2009

# A NEW LATENT CURE RATE MARKER MODEL FOR SURVIVAL DATA

By Sungduk Kim, Yingmei Xi and Ming-Hui Chen[1]

*National Institute of Child Health and Human Development,
Biogen Idec Inc. and University of Connecticut*

To address an important risk classification issue that arises in clinical practice, we propose a new mixture model via latent cure rate markers for survival data with a cure fraction. In the proposed model, the latent cure rate markers are modeled via a multinomial logistic regression and patients who share the same cure rate are classified into the same risk group. Compared to available cure rate models, the proposed model fits better to data from a prostate cancer clinical trial. In addition, the proposed model can be used to determine the number of risk groups and to develop a predictive classification algorithm.

**1. Introduction.** In cure rate modeling of event-time data, a fraction of the population is considered to have zero hazard. The model is often suitable for survival data from cancer clinical trials owing to advances in medical treatment and health care. For example, treatment of prostate cancer routinely cures the patient in the sense of completely eradicating the disease. Existing cure rate models are able to accommodate a fraction of the population being cured [Berkson and Gage (1952) and Maller and Zhou (1996)] and characteristics of tumor growth [Chen, Ibrahim and Sinha (1999), Tsodikov, Ibrahim and Yakovlev (2003) and Cooner et al. (2007)]. However, risk-group information is not easily incorporated. Prostate cancer patients can be classified into low, intermediate and high-risk groups on the basis of pre-treatment characteristics, such as the level of prostate-specific antigen (PSA), biopsy Gleason scores or clinical tumor categories [D'Amico et al. (1998, 2002)]. A failure to incorporate risk stratification into the cure rate model can lead to

Received September 2007; revised February 2009.
[1]Dr. Chen's research was supported in part by NIH Grants #CA 074015 and #GM 70335.

This is an electronic reprint of the original article published by the Institute of Mathematical Statistics in *The Annals of Applied Statistics*, 2009, Vol. 3, No. 3, 1124–1146. This reprint differs from the original in pagination and typographic detail.

poorly fitting statistical models and poorly estimated cure rates and predictive probabilities of risk groups. We address this problem via a latent class analysis of the cure rate model.

We consider data from a retrospective cohort study of $n = 1235$ men treated with radical prostatectomy (RP) at Brigham and Women's Hospital between 1988–2001, which is a subset of the data published in D'Amico et al. (2002). The primary endpoint is the time to prostate-specific antigen (PSA) recurrence or to the last follow-up, whichever came first. There were 261 patients who had PSA recurrence after the radical prostatectomy. We consider four prognostic factors: natural logarithm of prostate specific antigen (LogPSA) prior to RP, biopsy Gleason score, the 1992 American Joint Commission on Cancer (AJCC) clinical tumor category and the year of radical prostatectomy (Year). D'Amico et al. (2002) considered three risk groups based on PSA, biopsy Gleason score and clinical tumor category and reported the estimates of 8-year PSA recurrence free survival for the three risk groups based on the Kaplan–Mier (KM) method [Kaplan and Meier (1958)]. For the subset of the data considered here, the KM estimates of 8-year PSA recurrence free survival are 88.7%, 57.4% and 23.4% for low-risk patients (T1c, T2a, a PSA level $\leq 10$ ng/mL), intermediate-risk patients (T2b or Gleason score 7 or a PSA level $> 10$ and $\leq 20$ ng/mL) and high-risk patients (T2c or PSA level $> 20$ ng/mL or Gleason score $\geq 8$), respectively. This risk classification does capture that the low-risk patients have the highest PSA recurrence free survival and the high-risk patients have the lowest PSA recurrence free survival. However, there are some limitations of this risk classification. First, this risk classification is deterministic. In other words, it is not associated with predictive probabilities of risk groups. Second, it may be problematic, especially for those patients whose clinical characteristics fall within the boundary between two risk groups. Third, this risk classification is not flexible enough to incorporate other potentially important risk factors. For example, the year of diagnosis or treatment may have a significant effect on the "cure rate" and, thus, it may be an important factor for risk classification. To overcome these limitations, we develop a predictive classification algorithm based on the latent cure rate marker model. This algorithm first computes the probabilities of risk groups for a patient based on his clinical characteristics and then classifies him to a particular risk group with the largest predictive probability. As shown in Table 4 in Section 5, for a patient who had PSA of 5, and Gleason score $\leq 6$, clinical stage T1 and surgery in 2001, the predictive probabilities for three risk groups are 0.745, 0.241 and 0.014 and, thus, this patient will be classified into the "low risk" group.

Overdiagnosis of clinically insignificant prostate cancer was considered a major issue of prostate-specific antigen (PSA) screening since the U.S. Food and Drug Administration approved PSA testing in 1986 as a way to monitor prostate cancer progression [Wang and Arnold (2002)]. Etzioni et al. (2002) estimated rates of prostate cancer overdiagnosis due to PSA testing among men who were 60 to 84 years old in 1988. Overdiagnosis may occur when older men or men with comorbid illness who have very low risk disease are treated. However, overdiagnosis is usually not the case for men treated with surgery because they are healthy but they can have very low risk disease. Since the data we analyze in this paper were from those men who went to surgery, it may be appropriate to fit a cure rate model to this particular prostate cancer data set. We include the year of RP in the analysis, as it may have a significant effect on the "cure rate." There are two reasons for this. With time people are diagnosed after several PSA tests and serial screened men are diagnosed earlier with more favorable disease [e.g., Martin et al. (2008)] and with increased medical experience. Over time, the techniques of treatment also improve, which can improve outcome due to a learning curve, especially when new surgery (e.g., robotic RP) or radiation therapy (e.g., seed therapy) techniques are used. We fit both the Cox proportional hazards regression model [Cox (1972)] and the proposed latent cure rate marker model with a piecewise exponential baseline hazard function to this prostate cancer data. We then computed the logarithm of pseudomarginal likelihood (LPML) [Ibrahim, Chen and Sinha (2001), Chapter 6] and the Deviance Information Criterion (DIC) proposed by Spiegelhalter et al. (2002) for each model. From Table 2 in Section 5, we see that the best LPML and DIC values are $-821.5$ and $1640.8$ for the Cox model and $-816.0$ and $1613.7$ for the latent cure rate marker model. These results indicate that the cure rate model fit the data much better than the noncure rate model. Thus, a cure rate model is indeed needed for this data set.

Section 2 provides the detailed development of the proposed latent cure rate model. The prior and posterior are discussed in Section 3. The posterior predictive classification algorithm is developed in Section 4. Section 5 presents an analysis of the prostate cancer data. We conclude the paper with brief discussions in Section 6.

## 2. The models.

2.1. *Preliminary.* Let $y_i$ denote the observed survival time and let $\nu_i$ be the censoring indicator that equals 1 if $y_i$ is a failure time and 0 if it is right censored for the $i$th subject. Also, let $N_i$ denote the number of metastatic-competent tumor cells and assume that the $N_i$'s are independent Poisson random variables with mean $\theta_i$. Suppose further that $W_{ij}$ denotes the random time for the $j$th carcinogenic cell to produce a detectable cancer mass (incubation time for the $j$th carcinogenic cell) for the $i$th subject. We assume that the variables $W_{ij}$, $i = 1, 2, \ldots$, are independent and distributed with a common distribution function $F(y)$, and are independent of $N_i$. The time to relapse of cancer can be defined by the random variable $Y_i = \min\{W_{ij}, 0 \leq j \leq N_i\}$, where $P(W_{i0} = \infty) = 1$. Then, the survival function for the cure rate model for the $i$th subject is given by

$$(1) \qquad S_i(y) = P(Y_i \geq y) = \exp\{-\theta_i F(y)\}.$$

Using (1), the cure rate is given by $S_i(\infty) = \exp(-\theta_i)$, which is also equal to $P(N_i = 0)$. Other properties of the cure rate model (1) can be found in Yakovlev et al. (1993), Yakovlev and Tsodikov (1996) and Chen, Ibrahim and Sinha (1999). To build a regression model, Chen, Ibrahim and Sinha (1999) introduced covariates through $\theta_i$ via $\theta_i \equiv \theta(\mathbf{x}_i'\boldsymbol{\beta}) = \exp(\mathbf{x}_i'\boldsymbol{\beta})$, where $\mathbf{x}_i = (x_{i1}, x_{i2}, \ldots, x_{ip})'$ denotes the $p \times 1$ vector of covariates for the $i$th subject and $\boldsymbol{\beta} = (\beta_1, \beta_2, \ldots, \beta_p)'$ is the corresponding vector of regression coefficients, $i = 1, 2, \ldots, n$. Let $S(y) = 1 - F(y)$ and $f(y) = \frac{d}{dy}F(y)$. Then the resulting survival function is given by

$$(2) \qquad S_i(y|\mathbf{x}_i, \boldsymbol{\beta}) = \exp\{-\exp(\mathbf{x}_i'\boldsymbol{\beta})F(y)\}.$$

We refer to (2) as the CIS model.

A natural extension of the CIS model is the cure rate double regression model. Let $\boldsymbol{\beta}_1 = (\beta_{11}, \beta_{12}, \ldots, \beta_{1p})'$ and $\boldsymbol{\beta}_2 = (\beta_{21}, \beta_{22}, \ldots, \beta_{2p})'$. We assume $\theta_i = \exp(\mathbf{x}_i'\boldsymbol{\beta}_1)$. A proportional hazards model is assumed for the distribution function $F(y)$ of the incubation time for noncured subjects. Specifically, let the cumulative hazard function $H(y) = \exp(\mathbf{x}_i'\boldsymbol{\beta}_2)H_0(y)$, where $H_0(y)$ is the baseline cumulative hazard function. Then, $F(y) = 1 - \exp\{-H(y)\} = 1 - \exp\{-\exp(\mathbf{x}_i'\boldsymbol{\beta}_2)H_0(y)\}$. Under this assumption for $F(y)$, we have

$$(3) \qquad S_i(y|\mathbf{x}_i, \boldsymbol{\beta}) = \exp(-\exp(\mathbf{x}_i'\boldsymbol{\beta}_1)[1 - \exp\{-\exp(\mathbf{x}_i'\boldsymbol{\beta}_2)H_0(y)\}]),$$

where $\boldsymbol{\beta} = (\boldsymbol{\beta}_1', \boldsymbol{\beta}_2')'$. Yakovlev and Tsodikov (1996) used parametric accelerated failure time effects on the cumulative hazards with a similar idea. Model (3) in its semiparametric form has appeared in Broet et al. (2001), where they tended to use a generalized Gompertz name for the model. We see, from (3), that the model in (3) incorporates the covariates into both the cure rate and the hazard function with double proportional hazards structures. Thus, we refer to this model as the PHPH model. The name "PHPH" was also introduced by Tsodikov (2002) for the semiparametric version of the model. Recently, Liu, Lu and Shao (2006) developed the PHPH version of the standard cure rate model of Berkson and Gage (1952).

Another extension of the CIS model is the latent activation cure rate (LACR) model proposed by Cooner et al. (2007). Given $N_i \geq 1$, let $W_{i(1)} \leq W_{i(2)} \leq \cdots \leq W_{i(N_i)}$ denote the ordered values of the $W_{ij}$'s. The time to relapse of cancer is defined by $Y_i = W_{i(R_i)}$ for $1 \leq R_i \leq N_i$ and $W_{i0}$ if $N_i = 0$, where $R_i$ is an integer valued variable. Cooner et al. (2007) specified a conditional distribution for $R_i$ given $N_i$, denoted by $[R_i|N_i]$. When $N_i$ follows a Poisson distribution with mean $\theta_i = \exp(\mathbf{x}_i\boldsymbol{\beta})$ and $[R_i|N_i]$ is a discrete uniform on $\{1, 2, \ldots, N_i\}$ with probability $\frac{1}{N_i}$, the survival function under the LACR model is given by

$$(4) \qquad S_i(y|\mathbf{x}_i, \boldsymbol{\beta}) = \exp\{-\exp(\mathbf{x}_i\boldsymbol{\beta})\} + [1 - \exp\{-\exp(\mathbf{x}_i\boldsymbol{\beta})\}]S(y).$$

Other distributions for $N_i$ and $[R_i|N_i]$ are also considered in Cooner et al. (2007).

### 2.2. A new latent cure rate marker model.

The latent cure rate marker (LCRM) model assumes that the $N_i$'s are independent Poisson random variables with mean $\theta_{g_i}$, where $g_i$ is a (unknown) group indicator, and $\exp(-\theta_{g_i})$ is the cure rate marker. Let $G$ denote the number of distinct values of $\theta_{g_i}$. Further, $g_i$ $(1 \le g_i \le G)$ indicates the group membership. Without loss of generality, we assume $\theta_1 < \theta_2 < \cdots < \theta_G$. Under these constraints, the group membership $g_i$ is uniquely defined. Similar to the PHPH model, we assume the proportional hazards model for cumulative hazard function $H(y)$, that is, $H(y) = \exp(\mathbf{x}_i'\boldsymbol{\beta})H_0(y)$, where $H_0(y)$ is the baseline cumulative hazard function. Then, under the LCRM model, the conditional survival function of $y_i$ given $\theta_{g_i}$ is of the form

$$(5) \qquad S_i(y|\mathbf{x}_i, \boldsymbol{\beta}, \theta_{g_i}) = \exp(-\theta_{g_i}[1 - \exp\{-\exp(\mathbf{x}_i'\boldsymbol{\beta})H_0(y)\}]).$$

We assume a multinomial logistic regression model for the latent group membership $g_i$. To this end, let $\mathbf{z}_i' = (z_{i0}, z_{i1}, z_{i2}, \ldots, z_{iq})$ denote a $(q+1) \times 1$ vector of covariates for the $i$th subject, which includes an intercept (i.e., $z_{i0} = 1$) for $i = 1, 2, \ldots, n$. Also let $\boldsymbol{\phi}_j = (\phi_{j0}, \phi_{j1}, \phi_{j2}, \ldots, \phi_{jq})'$ denote the corresponding vectors of regression coefficients for $j = 1, 2, \ldots, G$ and $\boldsymbol{\phi}' = (\boldsymbol{\phi}_1', \boldsymbol{\phi}_2', \ldots, \boldsymbol{\phi}_{G-1}')$. Then the density of the group membership $g_i$ is given by

$$(6) \qquad f(g_i|\mathbf{z}_i, \boldsymbol{\phi}) = \frac{\exp(\mathbf{z}_i'\boldsymbol{\phi}_{g_i})}{\sum_{l=1}^{G} \exp(\mathbf{z}_i'\boldsymbol{\phi}_l)}.$$

For notational convenience, we let $\boldsymbol{\phi}_G = (0, 0, \ldots, 0)'$. Write $\boldsymbol{\theta}' = (\theta_1, \theta_2, \ldots, \theta_G)$. Using (5), the unconditional survival function of $y_i$ is given by

$$
\begin{aligned}
(7) \quad & S_i(y|\mathbf{x}_i, \mathbf{z}_i, \boldsymbol{\beta}, \boldsymbol{\theta}, \boldsymbol{\phi}) \\
& = \sum_{k=1}^{G} \exp(-\theta_k[1 - \exp\{-\exp(\mathbf{x}_i'\boldsymbol{\beta})H_0(y)\}]) \frac{\exp(\mathbf{z}_i'\boldsymbol{\phi}_k)}{\sum_{l=1}^{G} \exp(\mathbf{z}_i'\boldsymbol{\phi}_l)}.
\end{aligned}
$$

Unlike the CIS model, the LCRM model does not directly link the covariates to the cure fractions and instead it assumes that the population is characterized by an unobserved cure rate marker, namely, $\exp(-\theta_{g_i})$, where the latent group membership $g_i$ is described according to covariates via a multinomial logistic regression model. We note that the monotonic constraints on the cure rates $\theta_k$'s not only define the group membership $g_i$ but also ensure identifiability of the multinomial logistic regression model. We also see, from (7), that the LCRM model is indeed a finite mixture of cure rate models. If $\theta_k \to \theta$ for $k = 1, 2, \ldots, G$, (7) reduces to

$$S_i(y|\mathbf{x}_i, \boldsymbol{\beta}, \theta) = \exp(-\theta[1 - \exp\{-\exp(\mathbf{x}_i'\boldsymbol{\beta})H_0(y)\}]),$$

which is a special case of the PHPH model.

We assume the piecewise exponential model for the baseline hazard function $h_0(y)$, which is constructed as follows. We first partition the time axis into $J$ intervals: $(s_0, s_1]$, $(s_1, s_2]$, \ldots, $(s_{J-1}, s_J]$, where $s_0 = 0 < s_1 < s_2 < \cdots < s_J = \infty$. We then assume a constant hazard $\lambda_j$ over the $j$th interval $I_j = (s_{j-1}, s_j]$. That is, $h_0(y) = \lambda_j$ if $y \in I_j$ for $j = 1, 2, \ldots, J$. Then the corresponding cumulative distribution function (c.d.f.), $F_0(y|\boldsymbol{\lambda})$, is given by

$$(8) \qquad F_0(y|\boldsymbol{\lambda}) = 1 - \exp\left\{-\lambda_j(y - s_{j-1}) - \sum_{g=1}^{j-1} \lambda_g(s_g - s_{g-1})\right\}$$

for $s_{j-1} \leq y < s_j$, where $\boldsymbol{\lambda} = (\lambda_1, \ldots, \lambda_J)'$. We note that when $J = 1$, $F_0(y|\boldsymbol{\lambda})$ reduces to the parametric exponential model.

Let $D = (n, \mathbf{y}, X, Z, \boldsymbol{\nu}, \mathbf{N}, \mathbf{g})$ denote the complete data, where $\mathbf{y} = (y_1, \ldots, y_n)'$, $\boldsymbol{\nu} = (\nu_1, \nu_2, \ldots, \nu_n)'$, $X$ is the $n \times p$ matrix of covariates with $i$th row $\mathbf{x}_i'$, $Z$, which may share common components with $X$, is a $q$-vector of covariates with $i$th row $\mathbf{z}_i'$, $\mathbf{N}' = (N_1, N_2, \ldots, N_n)$, and $\mathbf{g}' = (g_1, g_2, \ldots, g_n)$. Then, the complete data likelihood under the LCRM model is given by

$$L(\boldsymbol{\beta}, \boldsymbol{\theta}, \boldsymbol{\phi}, \boldsymbol{\lambda}|D)$$

$$= \prod_{i=1}^{n} \left[ \prod_{j=1}^{J} (N_i \lambda_j)^{\nu_i \delta_{ij}} \right.$$

$$(9) \qquad \times \exp\left\{\nu_i \delta_{ij} \mathbf{x}_i'\boldsymbol{\beta} - \exp(\mathbf{x}_i'\boldsymbol{\beta}) N_i \delta_{ij} \right.$$

$$\left. \left. \times \left(\lambda_j(y_i - s_{j-1}) + \sum_{k=1}^{j-1} \lambda_k(s_k - s_{k-1})\right)\right\}\right]$$

$$\times \exp\left[\sum_{i=1}^{n}\left\{N_i \log \theta_{g_i} - \log(N_i!) - \theta_{g_i} + \mathbf{z}_i'\boldsymbol{\phi}_{g_i} - \log\left[\sum_{l=1}^{G} \exp(\mathbf{z}_i'\boldsymbol{\phi}_l)\right]\right\}\right],$$

where $\delta_{ij} = 1$ if the $i$th subject failed or was censored in the $j$th interval $I_j$, and 0 otherwise. Since $\mathbf{N}$ and $\mathbf{g}$ are not observed, by summing (9) over **N** and **g**, we obtain the likelihood function based on the observed data $D_{obs} = (n, \mathbf{y}, X, Z, \boldsymbol{\nu})$ given by

$$
\begin{aligned}
L(\boldsymbol{\beta}, \boldsymbol{\theta}, \boldsymbol{\phi}, \boldsymbol{\lambda} | D_{obs}) \\
= \prod_{i=1}^{n} \sum_{k=1}^{G} \Bigg( &\exp\Bigg[ \nu_i \Bigg\{ \log \theta_k + \mathbf{x}_i' \boldsymbol{\beta} + \sum_{j=1}^{J} \delta_{ij} \log \lambda_j - \exp(\mathbf{x}_i' \boldsymbol{\beta}) H_0^*(y_i) \Bigg\} \\
& - \theta_k (1 - \exp\{ -\exp(\mathbf{x}_i' \boldsymbol{\beta}) H_0^*(y_i) \}) \Bigg] \\
& \times \exp\Bigg\{ \mathbf{z}_i' \boldsymbol{\phi}_k - \log\Bigg( \sum_{l=1}^{G} \exp(\mathbf{z}_i' \boldsymbol{\phi}_l) \Bigg) \Bigg\} \Bigg),
\end{aligned}
\tag{10}
$$

where $H_0^*(y_i) = \sum_{j=1}^{J} \delta_{ij} [\lambda_j (y_i - s_{j-1}) + \sum_{l=1}^{j-1} \lambda_l (s_l - s_{l-1})]$.

**3. The prior and posterior distributions under the LCRM model.** We consider a joint prior distribution for $(\boldsymbol{\beta}, \boldsymbol{\theta}, \boldsymbol{\phi}, \boldsymbol{\lambda})$. Suppose that $J$ and $s_j$, $j = 1, \ldots, J$, are fixed. First we consider a fixed $G$. We assume that $\boldsymbol{\beta}$, $\boldsymbol{\theta}$, $\boldsymbol{\phi}$ and $\boldsymbol{\lambda}$ are independent a priori. Thus, the joint prior for $(\boldsymbol{\beta}, \boldsymbol{\theta}, \boldsymbol{\phi}, \boldsymbol{\lambda})$ is of the form $\pi(\boldsymbol{\beta}, \boldsymbol{\theta}, \boldsymbol{\phi}, \boldsymbol{\lambda}) = \pi(\boldsymbol{\beta})\pi(\boldsymbol{\theta})\pi(\boldsymbol{\phi})\pi(\boldsymbol{\lambda})$. We further assume that

$$
\boldsymbol{\beta} \sim N_p(0, c_{01} I_p), \qquad \boldsymbol{\phi}_k \sim N_q(0, c_{02} I_q), \qquad k = 1, 2, \ldots, G-1,
\tag{11}
$$

$$
\pi(\boldsymbol{\lambda}) \propto \prod_{j=1}^{J} \lambda_j^{a_0 - 1} \exp(-b_0 \lambda_j),
\tag{12}
$$

and

$$
\pi(\boldsymbol{\theta}) \propto \prod_{k=1}^{G} \theta_k^{a_k - 1} \exp(-b_k \theta_k), \qquad 0 < \theta_1 < \theta_2 < \cdots < \theta_G,
\tag{13}
$$

where $c_{01}$, $c_{02}$, $a_0$, $b_0$, $a_k$ and $b_k$, $k = 1, 2, \ldots, G$, are the prespecified hyperparameters. Due to the monotonic constraints, $\theta_1 < \theta_2 < \cdots < \theta_G$, eliciting the hyper-parameters $a_k$ and $b_k$ becomes more crucial than other hyperparameters. To this end, we first specify $\boldsymbol{\theta}_0 = (\theta_{01}, \theta_{02}, \ldots, \theta_{0G})'$ such that $\theta_{01} < \theta_{02} < \cdots < \theta_{0G}$. Equivalently, we specify a set of prior cure rates $\exp(-\theta_{0k})$, $k = 1, 2, \ldots, G$. Then, we set $a_k = \frac{1}{c_0^2}$ and $b_k = \frac{1}{c_0^2 \theta_{0k}}$, where $c_0$ is a known constant. This essentially implies that we specify the prior mean and the prior standard deviation of $\theta_k$ to be $\theta_{0k}$ and $c_0 \theta_{0k}$. Thus, $c_0$ quantifies the prior uncertainty in $\theta_{0k}$. A large value of $c_0$ reflects a vague prior belief in $\theta_{0k}$ and a small value of $c_0$ yields a strong prior belief in $\theta_{0k}$. In Section 5 we use the LPML and DIC measures to guide the choice of $c_0$ and $\boldsymbol{\theta}_0$ and we then conduct a sensitivity analysis on various choices of $c_0$ and $\boldsymbol{\theta}_0$.

Based on the prior distributions specified above, the joint posterior distribution of $\boldsymbol{\beta}$, $\boldsymbol{\theta}$, $\boldsymbol{\phi}$, $\boldsymbol{\lambda}$, $\mathbf{N}$ and $\mathbf{g}$ based on the complete data $D$ is thus given by

$$(14) \qquad \pi(\boldsymbol{\beta}, \boldsymbol{\theta}, \boldsymbol{\phi}, \boldsymbol{\lambda}, \mathbf{N}, \mathbf{g} | D_{obs}) \propto L(\boldsymbol{\beta}, \boldsymbol{\theta}, \boldsymbol{\phi}, \boldsymbol{\lambda} | D) \pi(\boldsymbol{\beta}) \pi(\boldsymbol{\theta}) \pi(\boldsymbol{\phi}) \pi(\boldsymbol{\lambda}),$$

where $L(\boldsymbol{\beta}, \boldsymbol{\theta}, \boldsymbol{\phi}, \boldsymbol{\lambda} | D)$ is defined in (9). We note that when the priors $\pi(\boldsymbol{\beta})$, $\pi(\boldsymbol{\theta})$, $\pi(\boldsymbol{\phi})$ and $\pi(\boldsymbol{\lambda})$ are proper, the resulting posterior is proper. However, even when we take an improper prior for $\boldsymbol{\theta}$, an improper uniform prior for $\boldsymbol{\beta}$ and an improper Jeffreys-type prior for $\boldsymbol{\lambda}$, that is, $c_{01} \to \infty$ and $a_0 = b_0 = 0$ in (12), the posterior is still proper under some mild conditions. We formally state this result in the following theorem.

THEOREM 1. *Suppose that $\pi(\boldsymbol{\beta}) \propto 1$ and $\pi(\boldsymbol{\lambda}) \propto \prod_{j=1}^{J} \lambda_j^{-1}$. Let $X_j$ be an $n \times (p+1)$ matrix with its $i$th row equal to $\nu_i \delta_{ij}(1, x_i')$, where $p$ is the dimension of $\boldsymbol{\beta}$. Assume that* (i) *when $\nu_i = 1$, $y_i > 0$,* (ii) *$d_j \equiv \sum_{i=1}^{n} \nu_i \delta_{ij} \geq 1$ for $j = 1, \ldots, J$,* (iii) *there exists a $j^*$ such that $X_{j^*}$ is of full rank, and* (iv) *$c_{02} > 0$, $a_k > 0$ and $b_k \geq 0$ for $k = 1, 2, \ldots, G-1$, $d + \sum_{k=1}^{G-1} a_k + a_G > 0$, where $d = \sum_{j=1}^{J} d_j$, and $b_G > 0$. Then, the resulting posterior in (14) is proper.*

The proof of the theorem is given in the Appendix. The conditions (i)–(iii) are indeed quite mild and essentially require that all event times are strictly positive, at least one event occurs in each chosen interval $(s_{j-1}, s_j]$, and the covariate matrix is of full rank for at least one interval. These conditions are easily satisfied in most applications and are quite easy-to-check. We note that the condition (iv) does not require $a_G > 0$. Thus, $\pi(\boldsymbol{\theta})$ can be improper. We also note that the latent structure of the LCRM model leads to the development of a Markov chain Monte Carlo (MCMC) algorithm for sampling from posterior distribution in (14). When $G$ is not specified, we assume a truncated Poisson distribution with mean $\mu_G$ on $\{1, 2, \ldots, G_{\max}\}$ for $G$, where $\mu_G$ and $G_{\max}$ are prespecified. Then, we develop a reversible jump algorithm for carrying out posterior computation. The description of the MCMC algorithm for a fixed $G$ and the detailed development of the reversible jump MCMC based on Lopes and West (2004) and Green (1995) are given in online supplementary material [Kim, Xi and Chen (2009)].

**4. Posterior predictive classification under the LCRM model.** In this section we consider classification via the posterior predictive probability. The latent cure rate markers under the LCRM model can be naturally used for the predictive classification. Let $\mathbf{x}_{new}$ and $\mathbf{z}_{new}$ denote the future values of the vectors of baseline covariates. Also let $g_{new}$ denote the future group indicator. Suppose that $g_{new}$ takes a value between 1 and $G$, where $G$ is fixed. Then, the conditional posterior probability for $g_{new}$ given $\boldsymbol{\phi}$ and $\mathbf{z}_{new}$ is given by

$$(15) \quad P(g_{new} = k|\boldsymbol{\phi}, \mathbf{z}_{new}, G) = \frac{\exp(\mathbf{z}'_{new}\boldsymbol{\phi}_k)}{\sum_{l=1}^{G} \exp(\mathbf{z}'_{new}\boldsymbol{\phi}_l)}, \qquad k = 1, 2, \ldots, G.$$

The posterior estimate of this predictive probability for $g_{new}$ is the posterior expectation of $P(g_{new} = k|\boldsymbol{\phi}, \mathbf{z}_{new})$ given by

$$(16) \quad \hat{p}(k|\mathbf{z}_{new}, G) = E[P(g_{new} = k|\boldsymbol{\phi}, \mathbf{z}_{new}, G)|D_{obs}]$$

for $k = 1, 2, \ldots, G$, where the expectation is taken with respect to the posterior distribution of $\boldsymbol{\phi}$ based on the observed data $D_{obs}$. The clinical interpretation of (16) is that, given the patient's characteristic $\mathbf{z}_{new}$, $\hat{p}(k|\mathbf{z}_{new})$ is the probability that the patient is in the $k$th risk group.

Next, we consider the conditional predictive probability for $g_{new} = k$ given the survival time $Y \geq t$, $\mathbf{x}_{new}$ and $\mathbf{z}_{new}$. This conditional predictive probability can be calculated as follows:

$$(17) \quad \begin{aligned} &P(g_{new} = k|\boldsymbol{\beta}, \boldsymbol{\theta}, \boldsymbol{\phi}, \boldsymbol{\lambda}, t, \mathbf{x}_{new}, \mathbf{z}_{new}, G) \\ &= \frac{\exp(\mathbf{z}'_{new}\boldsymbol{\phi}_k - \theta_k[1 - \exp\{-\exp(\mathbf{x}'_{new}\boldsymbol{\beta})H_0^*(t)\}])}{\sum_{l=1}^{G} \exp(\mathbf{z}'_{new}\boldsymbol{\phi}_l - \theta_l[1 - \exp\{-\exp(\mathbf{x}'_{new}\boldsymbol{\beta})H_0^*(t)\}])}, \end{aligned}$$

where $H_0^*(t)$ is given in (10). The posterior estimate of (17) for $g_{new}$ is

$$(18) \quad \hat{p}(k|t, \mathbf{x}_{new}, \mathbf{z}_{new}, G) = E[P(g_{new} = k|\boldsymbol{\beta}, \boldsymbol{\theta}, \boldsymbol{\phi}, \boldsymbol{\lambda}, t, \mathbf{x}_{new}, \mathbf{z}_{new})|D_{obs}].$$

From (16) and (18), it is easy to see that

$$\hat{p}(k|\mathbf{z}_{new}, G) = \hat{p}(k|t = 0, \mathbf{x}_{new}, \mathbf{z}_{new}, G)$$

for $k = 1, 2, \ldots, G$. Since $\lim_{t \to \infty} H_0^*(t) = \infty$, we also have

$$(19) \quad \begin{aligned} \lim_{t \to \infty} \hat{p}(k|t, \mathbf{x}_{new}, \mathbf{z}_{new}, G) &= E\left[\frac{\exp(\mathbf{z}'_{new}\boldsymbol{\phi}_k - \theta_k)}{\sum_{l=1}^{G} \exp(\mathbf{z}'_{new}\boldsymbol{\phi}_l - \theta_l)}\right], \\ &\qquad\qquad k = 1, 2, \ldots, G. \end{aligned}$$

Using the posterior predictive probability in (18), we classify a new patient with characteristic $(\mathbf{x}_{new}, \mathbf{z}_{new})$ into risk group $k^*$ if

$$(20) \quad k^* = \underset{1 \leq k \leq G}{\arg\max} \ \hat{p}(k|t, \mathbf{x}_{new}, \mathbf{z}_{new}, G).$$

An attractive property of the posterior predictive probability in (18) is presented in the following theorem.

THEOREM 2. *The posterior predictive probability for the lowest risk group* ($k = 1$), $\hat{p}(1|t, \mathbf{x}_{new}, \mathbf{z}_{new}, G)$, *increases in* $t$, *while for the highest risk group* ($k = G$), $\hat{p}(G|t, \mathbf{x}_{new}, \mathbf{z}_{new}, G)$ *decreases in* $t$.

The proof of Theorem 2 is given in the Appendix. Based on (19) and Theorem 2, we have that, for $t \geq 0$,

$$(21) \qquad \hat{P}(g_{new} = 1 | t, \mathbf{x}_{new}, \mathbf{z}_{new}, G) \leq E\left[\frac{\exp(\mathbf{z}'_{new}\boldsymbol{\phi}_1 - \theta_1)}{\sum_{k=1}^{G}\exp(\mathbf{z}'_{new}\boldsymbol{\phi}_k - \theta_k)}\right],$$

which is the largest probability that $\hat{P}(g_{new} = 1 | t, \mathbf{x}_{new}, \mathbf{z}_{new}, G)$ may achieve for $t > 0$ given the patient's characteristic $(\mathbf{x}_{new}, \mathbf{z}_{new})$. Similarly, we have

$$(22) \qquad \hat{P}(g_{new} = G | t, \mathbf{x}_{new}, \mathbf{z}_{new}, G) \geq E\left[\frac{\exp(\mathbf{z}'_{new}\boldsymbol{\phi}_G - \theta_G)}{\sum_{k=1}^{G}\exp(\mathbf{z}'_{new}\boldsymbol{\phi}_k - \theta_k)}\right],$$

which is the smallest probability that $\hat{P}(g_{new} = G | t, \mathbf{x}_{new}, \mathbf{z}_{new}, G)$ can get for $t > 0$. The quantity $\hat{P}(g_{new} = k | t, \mathbf{x}_{new}, \mathbf{z}_{new}, G)$ is clinically important as this gives the patient an idea how well he can do prospectively given his baseline characteristic.

Finally, we note that when $G$ is not specified, a similar posterior predictive classification algorithm can be established. For example, instead of (15), we compute

$$P(g_{new} = k | \boldsymbol{\phi}, \mathbf{z}_{new}) = \sum_{G=1}^{G_{\max}} \pi(G)\frac{\exp(\mathbf{z}'_{new}\boldsymbol{\phi}_k)}{\sum_{l=1}^{G}\exp(\mathbf{z}'_{new}\boldsymbol{\phi}_l)}, \qquad k = 1, 2, \ldots, G_{\max},$$

where $\pi(G)$ denotes the prior distribution for $G$ and $G_{\max}$ is the largest value of $G$.

## 5. Analysis of the prostate cancer data.

We revisit the prostate cancer data discussed in Section 1. The response variable $y$ is the time to prostate-specific antigen (PSA) recurrence. Covariates $x_1$, $x_2$, $x_3$, $x_4$ and $x_5$ correspond to LogPSA, biopsy Gleason score, clinical tumor category and the year of radical prostatectomy (Year). A summary of covariates is given in Table 1. The covariates LogPSA ($x_1$) and Year ($x_5$) are continuous, while $x_2$, $x_3$ and $x_4$ are binary. The mean and the standard deviation for LogPSA were 1.95 and 0.72. We also set $z_j = x_j$ for $j = 1, \ldots, 5$. In all computations we standardized all covariates by subtracting their respective sample means and then being divided by their respective sample standard deviations.

The hyper-parameters of the prior distribution in Section 4 are specified as follows. In (11), (12) and (13), we take $c_{01} = 1000$, $c_{02} = 3$, $a_0 = 1$, $b_0 = 0.01$ and $c_0 = 2.5$. We choose $c_{01}$ to be much larger than $c_{02}$ as the posterior is proper even when $\pi(\boldsymbol{\beta}) \propto 1$ according to Theorem 1. Also, $a_0 = 1$ and $b_0 = 0.01$ are specified so that the prior for $\boldsymbol{\lambda}$ is relatively noninformative. We further specify $\theta_{01} = -\log(0.5)$ for $G = 1$; $\theta_{01} = -\log(0.9)$ and $\theta_{02} = -\log(0.3)$ for $G = 2$; $\theta_{01} = -\log(0.9)$, $\theta_{02} = -\log(0.5)$ and $\theta_{03} = -\log(0.1)$ for $G = 3$; $\theta_{01} = -\log(0.9)$, $\theta_{02} = -\log(0.6)$, $\theta_{03} = -\log(0.3)$ and $\theta_{04} =$ $-\log(0.1)$ for $G = 4$; and $\theta_{01} = -\log(0.9)$, $\theta_{02} = -\log(0.7)$, $\theta_{03} = -\log(0.5)$, $\theta_{04} = -\log(0.3)$ and $\theta_{05} = -\log(0.1)$ for $G = 5$. We note that $(0.9, 0.5, 0.1)$ for $G = 3$ were determined by the KM estimates of cure rates based on the three risk groups defined in D'Amico et al. (1998, 2002).

Table 2 shows the values of LPML and DIC for the Cox, CIS, PHPH, LACR and LCRM models for various $J$'s and $G$'s. We note that under the Cox model, the survival function is given by $S_i(y|\mathbf{x}_i, \boldsymbol{\beta}, \boldsymbol{\lambda}) = \exp\{-\exp(\mathbf{x}_i'\boldsymbol{\beta})H_0(y|\boldsymbol{\lambda})\}$, where $H_0(y|\boldsymbol{\lambda})$ is the cumulative baseline hazard function corresponding to $F_0(y|\boldsymbol{\lambda})$ given in (8). From Table 2, we observe that there is a concave pattern in the LPMLs and there is a convex pattern in the DICs as functions of $J$ for the CIS, PHPH and LCRM with fixed $G$. We note that for $J = 15$, the values of LPML and DIC are $-827.9$ and 1648.9 for the Cox model and $-838.0$ and 1673.5 for the LACR model. Thus, the concave (convex) pattern in the LPMLs (DICs) as functions of $J$ still holds for the Cox and LCRM models. Similar patterns are also observed in the LPMLs and DICs as functions of $G$ for fixed $J$ under the LCRM. Among the three $J$'s shown in Table 2, $J = 5$ consistently fits the data better for the CIS, PHPH and LCRM models and $J = 10$ fits the data better for the Cox and LACR models. The LCRM model, with $J = 5$ and $G = 3$ fits the data best among all models considered. In particular, LPML $= -816.0$ and DIC $= 1613.7$ for the best LCRM model while LPML $= -821.5$ and DIC $= 1640.8$ for the best Cox model. Except for $G = 1$, the LCRM model with $G \geq 2$ improves the fit compared to the CIS, PHPH and LACR models.

The posterior estimates of the parameters under the best LCRM model with $J = 5$ and $G = 3$ are given in Table 3. We see from this table that LogPSA is significant in the proportional hazards model (5) for the survival function for a "noncured" subject and LogPSA, G8H and Year of RP are significant in the multinomial model (6) for the latent group membership at a significance level of 0.05. In addition, Year of RP is nearly significant in both models. Although the prior cure rates for the three risk groups are 0.9, 0.5 and 0.1, respectively, the resulting posterior estimates of these cure

TABLE 1
*Summary of covariates for prostate cancer data*

| Covariate | Coded variable | Value | Definition | Frequency |
|---|---|---|---|---|
| $x_1$ | LogPSA | $(-\infty, \infty)$ | Logarithm of PSA prior to RP | – |
| $(x_2, x_3)$ | (G7, G8H) | $(0, 0)$ | Gleason score 6 or less | 866 |
| | | $(1, 0)$ | Gleason score 7 | 303 |
| | | $(0, 1)$ | Gleason score 8–10 | 66 |
| $x_4$ | Cstage | 0 (T1) | Clinical tumor category T1c or T2a | 1055 |
| | | 1 (T2) | Clinical tumor category T2b or T2c | 180 |
| $x_5$ | Year | $>0$ | Year of RP | – |

TABLE 2

*LPMLs and DICs of Cox, CIS, PHPH, LACR and LCRM models*

| Model | G | J = 1 | | J = 5 | | J = 10 | |
|-------|---|-------|-----|-------|-----|--------|-----|
| | | LPML | DIC | LPML | DIC | LPML | DIC |
| Cox | | −864.0 | 1722.3 | −822.3 | 1642.4 | −821.5 | 1640.8 |
| CIS | | −827.4 | 1651.6 | −821.6 | 1641.5 | −822.4 | 1643.0 |
| PHPH | | −831.5 | 1655.5 | −824.2 | 1642.3 | −825.4 | 1646.9 |
| LACR | | −841.2 | 1680.6 | −832.2 | 1662.6 | −831.8 | 1661.6 |
| LCRM | 1 | −845.8 | 1686.5 | −821.3 | 1640.6 | −823.3 | 1645.4 |
| | 2 | −823.3 | 1633.5 | −820.8 | 1628.1 | −822.6 | 1632.3 |
| | 3 | −822.9 | 1626.3 | −816.0 | 1613.7 | −819.5 | 1624.6 |
| | 4 | −823.9 | 1628.4 | −817.1 | 1617.3 | −820.3 | 1627.4 |
| | 5 | −824.4 | 1629.1 | −818.0 | 1620.8 | −821.7 | 1634.9 |

TABLE 3

*Posterior estimates based on the best LCRM model*

| Variable | Posterior mean | Posterior SD | 95% HPD interval |
|----------|---------------|--------------|------------------|
| $\beta_1$ (LogPSA) | 0.349 | 0.107 | (0.136, 0.554) |
| $\beta_2$ (G7) | 0.117 | 0.135 | (−0.141, 0.395) |
| $\beta_3$ (G8H) | 0.090 | 0.085 | (−0.071, 0.260) |
| $\beta_4$ (Cstage) | 0.042 | 0.095 | (−0.138, 0.231) |
| $\beta_5$ (Year) | −0.269 | 0.143 | (−0.541, 0.016) |
| $\theta_1$ | 0.069 | 0.118 | (0.000, 0.301) |
| $\theta_2$ | 1.193 | 0.582 | (0.328, 2.316) |
| $\theta_3$ | 2.671 | 1.035 | (1.443, 4.490) |
| $\exp(-\theta_1)$ | 0.939 | 0.095 | (0.740, 1.000) |
| $\exp(-\theta_2)$ | 0.347 | 0.156 | (0.059, 0.625) |
| $\exp(-\theta_3)$ | 0.092 | 0.052 | (0.000, 0.181) |
| $\phi_{10}$ (Intercept) | 0.842 | 0.822 | (−0.907, 2.431) |
| $\phi_{11}$ (LogPSA) | −1.841 | 0.549 | (−2.962, −0.963) |
| $\phi_{12}$ (G7) | −1.162 | 0.924 | (−3.349, 0.183) |
| $\phi_{13}$ (G8H) | −1.801 | 1.068 | (−4.128, −0.030) |
| $\phi_{14}$ (Cstage) | −0.694 | 0.554 | (−1.818, 0.181) |
| $\phi_{15}$ (Year) | 0.840 | 0.373 | (0.106, 1.597) |
| $\phi_{20}$ (Intercept) | −0.385 | 1.152 | (−2.670, 1.827) |
| $\phi_{21}$ (LogPSA) | −3.123 | 0.987 | (−5.160, −1.127) |
| $\phi_{22}$ (G7) | 1.118 | 1.058 | (−0.991, 3.163) |
| $\phi_{23}$ (G8H) | −0.178 | 1.662 | (−3.479, 2.878) |
| $\phi_{24}$ (Cstage) | −0.741 | 1.266 | (−3.224, 1.628) |
| $\phi_{25}$ (Year) | 1.144 | 0.851 | (−0.519, 2.884) |

rates are 0.939, 0.347 and 0.092. Under the same model setting for Table 3, the posterior predictive probabilities, $\hat{p}(k|t, \mathbf{x}_{new}, \mathbf{z}_{new}, G)$ given in (19) with $\mathbf{z}_{new} = \mathbf{x}_{new}$, are computed for three sets of baseline covariates $\mathbf{x}_{new}$'s for various $t$'s and the results are given in Table 4. Based on the proposed classification criterion given in (20), these probabilities clearly indicate that a patient with a PSA level of 5, Gleason 6 or less, and tumor stage T1 belongs to risk group 1 (low risk group) and a patient with a PSA level of 30, Gleason 8 to 10, and tumor stage T2 falls into risk group 3 (high risk group) no matter whether he had surgery in 1988 or 2001. However, a patient with a PSA level of 5, Gleason 7 and tumor stage T2 may be classified into risk group 3 (high risk group) if he had surgery in 1988 while a patient with the same PSA level, Gleason score and tumor stage will be classified into risk group 2 (intermediate risk group) if he had surgery in 2001. From Table 4, we also see that for each set of baseline covariates, the risk classification does not change no matter how long the patient will live if he had surgery in 2001 and this is not the case when he had surgery in 1988. In addition, the overall cure rates,

$$S(\infty|\mathbf{x}_{new}, \mathbf{z}_{new}, \boldsymbol{\beta}, \boldsymbol{\theta}, \boldsymbol{\phi}) = \sum_{k=1}^{G} \exp(-\theta_k) \frac{\exp(\mathbf{z}'_{new}\boldsymbol{\phi}_k)}{\sum_{l=1}^{G} \exp(\mathbf{z}'_{new}\boldsymbol{\phi}_l)},$$

are presented in Table 4. It is interesting to see that when (PSA, Gleason, Cstage) = (5, ≤6, T1), the overall cure rate is much smaller than that given $g_{new} = 1$, when (PSA, Gleason, Cstage) = (5, 7, T2), the overall cure rate is greater than that given $g_{new} = 2$ if he had surgery in 2001, while the overall cure rate is very similar to the risk group specific cure rate ($g_{new} = 3$) when (PSA, Gleason, Cstage) = (30, 8–10, T2). Figure 1 shows the estimated risk group specific PSA recurrence free probabilities corresponding to these three sets of covariates and the estimated overall PSA recurrence free probability when the year of RP was 2001. From plots (a), (b) and (c), we see that three risk group specific probability curves are well separated from each other. These plots also show that a wrong classification may lead to either over-estimate or under-estimate of the PSA recurrence free probability. Thus, the posterior predictive classification is quite important, as a correct classification leads to more accurate estimates of the cure rate as well as the PSA recurrence free probability.

We further conducted a sensitivity analysis on the choice of $c_0$ and $\theta_{0j}$'s. Table 5 shows the LPML and DIC values of the LCRM model with $G = 3$ for various $c_0$ and the prior cure rates $\exp(\boldsymbol{\theta}_0) = (0.9, 0.5, 0.1)$, $(0.8, 0.5, 0.2)$ and $(0.7, 0.5, 0.3)$. Both LPML and DIC values are very similar for almost all choices of $c_0$. Among all values of $c_0$ and $\exp(\boldsymbol{\theta}_0)$, $c_0 = 2.5$ and $\exp(\boldsymbol{\theta}_0) = (0.9, 0.5, 0.1)$ yield the largest LPML and the smallest DIC among all choices considered. Although not reported in Table 5, the posterior estimates of the cure rates were also calculated under those choices of $c_0$ and the prior cure rates. For example, when $c_0 = 2.5$, the posterior estimates of the cure rates and the corresponding posterior standard deviations are (0.939, 0.347, 0.092) and (0.095, 0.156, 0.052) for $\exp(\boldsymbol{\theta}_0) = (0.9, 0.5, 0.1)$, (0.936, 0.351, 0.091) and (0.097, 0.161, 0.051) for $\exp(\boldsymbol{\theta}_0) = (0.8, 0.5, 0.2)$, and (0.936, 0.352, 0.094) and (0.087, 0.163, 0.051) for $\exp(\boldsymbol{\theta}_0) = (0.7, 0.5, 0.3)$. Similar results are obtained for other choices of $c_0$. These results demonstrate that the proposed LCRM model is quite robust to the specification of $c_0$ and prior cure rates.

TABLE 4
*Posterior predictive probability based on the best LCRM model*

| Year | PSA | Gleason | Stage | $t$ | $\hat{p}(k=1|t)$ | $\hat{p}(k=2|t)$ | $\hat{p}(k=3|t)$ | Overall cure rate |
|------|-----|---------|-------|-----|------------------|------------------|------------------|-------------------|
| 1988 | 5 | ≤6 | T1 | 0 | 0.692 | 0.099 | 0.209 | 0.705 |
| | | | | 5 | 0.814 | 0.085 | 0.101 | |
| | | | | ∞ | 0.910 | 0.061 | 0.029 | |
| | 5 | 7 | T2 | 0 | 0.057 | 0.384 | 0.559 | 0.234 |
| | | | | 5 | 0.132 | 0.412 | 0.456 | |
| | | | | ∞ | 0.238 | 0.426 | 0.336 | |
| | 30 | 8–10 | T2 | 0 | 0.001 | 0.053 | 0.947 | 0.098 |
| | | | | 5 | 0.005 | 0.068 | 0.928 | |
| | | | | ∞ | 0.008 | 0.072 | 0.920 | |
| 2001 | 5 | ≤6 | T1 | 0 | 0.745 | 0.241 | 0.014 | 0.781 |
| | | | | 5 | 0.770 | 0.220 | 0.010 | |
| | | | | ∞ | 0.868 | 0.130 | 0.002 | |
| | 5 | 7 | T2 | 0 | 0.183 | 0.657 | 0.160 | 0.409 |
| | | | | 5 | 0.220 | 0.653 | 0.127 | |
| | | | | ∞ | 0.327 | 0.618 | 0.055 | |
| | 30 | 8–10 | T2 | 0 | 0.006 | 0.143 | 0.851 | 0.117 |
| | | | | 5 | 0.020 | 0.166 | 0.815 | |
| | | | | ∞ | 0.042 | 0.186 | 0.772 | |

When $G$ is not specified, we used the RJMCMC algorithm given in Kim, Xi and Chen (2009). In the RJMCMC algorithm, we took $a = 3$ for $\theta_l$ and $d_l = 0.5$ for $\phi_l$, $l = 1, 2, \ldots, G - 1$. We specified the transition matrix as follows:

$$\mathrm{TR} = \begin{pmatrix} 0.0 & 1.0 & 0.0 & 0.0 & 0.0 \\ 0.5 & 0.0 & 0.5 & 0.0 & 0.0 \\ 0.0 & 0.5 & 0.0 & 0.5 & 0.0 \\ 0.0 & 0.0 & 0.5 & 0.0 & 0.5 \\ 0.0 & 0.0 & 0.0 & 1.0 & 0.0 \end{pmatrix}.$$

The dimension of model, $G$, is assumed to follow a Poisson distribution with mean $\mu_G = 3$ and truncated between 1 and 5. Also $J$ is fixed to

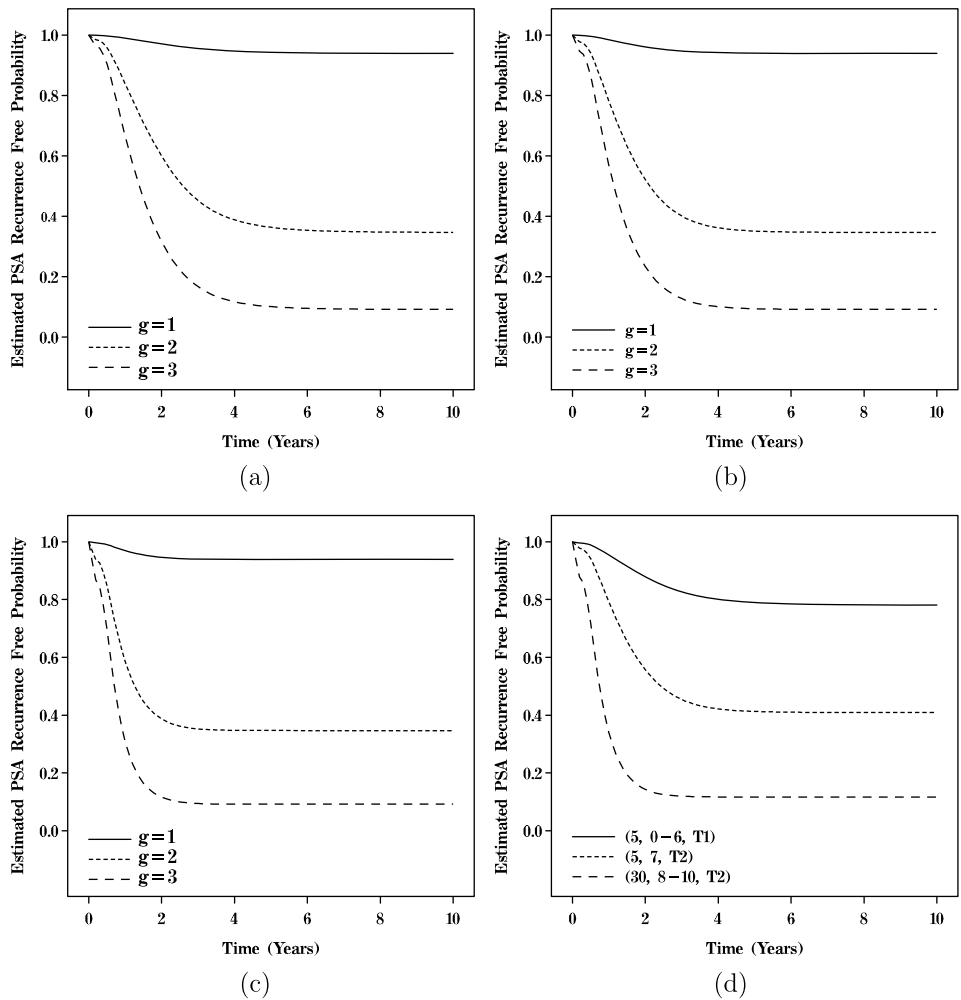

Fig. 1. *Plots of the estimated risk group specific PSA recurrence free probabilities corresponding to (PSA, Gleason, Cstage) = (5, ≤6, T1)* (a)*, (5, 7, T2)* (b)*, and (30, 8–10, T2)* (c)*, and the estimated overall PSA recurrence free probability* (d) *for year of RP = 2001.*

be 5. Under the above setting, the posterior probabilities of $G$ are computed and these are $P(G=1|D_{obs}) = 0.0$, $P(G=2|D_{obs}) = 0.224$, $P(G=3|D_{obs}) = 0.534$, $P(G=4|D_{obs}) = 0.242$, and $P(G=5|D_{obs}) = 0.0$, respectively. Therefore, the model with $G=3$ has the highest posterior model probability. This result is consistent with the best model identified by the LPML and DIC measures shown in Table 2. We also conducted a sensitivity analysis on the specification of $\mu_G$ in the prior distribution for $G$. Specifically, we obtained that $P(G=1|D_{obs}) = 0.0$, $P(G=2|D_{obs}) = 0.251$,

TABLE 5
*LPMLs and DICs of the LCRM model for various $c_0$ and prior cure rates*

| | **Prior cure rates** | | | | | |
| | **(0.9, 0.5, 0.1)** | | **(0.8, 0.5, 0.2)** | | **(0.7, 0.5, 0.3)** | |
| $c_0$ | **LPML** | **DIC** | **LPML** | **DIC** | **LPML** | **DIC** |
| 0.5 | −816.9 | 1615.4 | −817.7 | 1616.5 | −817.8 | 1617.2 |
| 1.0 | −816.6 | 1614.9 | −817.2 | 1616.0 | −817.2 | 1616.6 |
| 1.5 | −816.4 | 1614.3 | −816.7 | 1615.3 | −816.7 | 1616.0 |
| 2.0 | −816.2 | 1614.1 | −816.4 | 1615.0 | −816.6 | 1615.5 |
| 2.5 | −815.9 | 1613.7 | −816.2 | 1614.6 | −816.3 | 1615.1 |
| 3.0 | −816.1 | 1614.0 | −816.2 | 1614.8 | −816.4 | 1615.4 |
| 3.5 | −816.1 | 1614.4 | −816.3 | 1615.1 | −816.6 | 1615.8 |
| 4.0 | −816.4 | 1614.8 | −816.4 | 1615.5 | −817.0 | 1616.1 |
| 10.0 | −816.6 | 1615.4 | −816.9 | 1615.9 | −817.3 | 1616.5 |

$P(G = 3|D_{obs}) = 0.535$, $P(G = 4|D_{obs}) = 0.214$, and $P(G = 5|D_{obs}) = 0.0$ for $\mu_G = 2$, and $P(G = 1|D_{obs}) = 0.0$, $P(G = 2|D_{obs}) = 0.162$, $P(G = 3|D_{obs}) = 0.536$, $P(G = 4|D_{obs}) = 0.302$, and $P(G = 5|D_{obs}) = 0.0$ for $\mu_G = 4$. Thus, the model with $G = 3$ consistently has the highest posterior model probability for all three choices of $\mu_G$.

In all the computations, we first generated 100,000 Gibbs samples with a burn-in of 4000 iterations, and we then used 20,000 iterations obtained from every 5th iteration for computing all posterior estimates, including posterior mean, posterior standard deviation, 95% highest posterior density (HPD) intervals and LPML. The computer codes were written in FORTRAN 95 using IMSL subroutines with double precision accuracy. The convergence of the MCMC sampling algorithm was checked using several diagnostic procedures as recommended by Cowles and Carlin (1996).

**6. Discussions.** In Section 5 we used LPML and DIC measures to assess the goodness of fit of the models for different choices of $G$ and $J$. LPML is a well-established Bayesian model comparison criterion based on the conditional predictive ordinate (CPO) statistics, which is particularly suitable for the cure rate models. Let $\text{CPO}_i$ denote the CPO statistic for the $i$th subject. LPML is defined as

$$\text{LPML} = \sum_{i=1}^{n} \log(\text{CPO}_i).$$

The larger the LPML, the better the fit of a given model. Letting $\gamma$ denote the vector of all model parameters and $L(\gamma|D_{obs})$ the likelihood based on the observed data $D_{obs}$, the DIC is defined as

$$\text{DIC} = \text{Dev}(\bar{\gamma}) + 2p_D,$$

where $\mathrm{Dev}(\gamma) = -2 \log L(\gamma|D_{obs})$ is a deviance function, $\bar{\gamma}$ is the posterior mean of $\gamma$, $p_m = \overline{\mathrm{Dev}}(\gamma) - \mathrm{Dev}(\bar{\gamma})$, and $\overline{\mathrm{Dev}}(\gamma)$ is the posterior mean of $\mathrm{Dev}(\gamma)$. For the LCRM model, $\gamma = (\beta, \theta, \phi, \lambda)$ and $L(\gamma|D_{obs}) = L(\beta, \theta, \phi, \lambda|D_{obs})$, which is given by (10). The DIC is a Bayesian measure of predictive model performance, which is decomposed into a measure of fit and a measure of model complexity ($p_D$). The smaller the value of DIC, the better the model will predict new observations generated in the same way as the data. As discussed in Spiegelhalter et al. (2002), DIC is the Bayesian version of the Akaike Information Criterion (AIC) [Akaike (1973)]. Unlike AIC, the dimensional penalty in DIC is automatically calculated without actually counting the number of parameters. Although the dimensional penalty is not explicitly shown in LPML, LPML has a dimensional penalty similar to AIC as derived by Gelfand and Dey (1994) based on the asymptotic approximation. Moreover, as discussed in Ibrahim, Chen and Sinha (2001), the LPML measure is particularly suitable for comparing cure rate models, as the moments do not exist under these models.

As discussed in Sections 1 and 2, there are several cure rate models for survival data with a cure fraction recently developed in the literature. There is a distinct difference between the proposed model and the existing ones. Specifically, the new model is to no longer explain the cure fractions directly according to covariates but to divide the population into latent classes characterized by specific cure rates and being described according to covariates. This nice feature of the proposed model allows us to develop the predictive classification algorithm for classifying patients into different risk groups. The proposed mixture model falls within the latent class modeling framework. The latent class models are commonly used for analyzing complex sample survey data. For survey data, a latent class model is often used to explain unobservable categorical relationships or latent structures that characterize discrete multivariate data [Dayton (1999), Agresti (2002) and Patterson, Dayton and Graubard (2002)]. Recently, latent class models have been developed for survival data. Lin et al. (2002) proposed latent class models for joint longitudinal and survival data. They assumed a Cox proportional hazards model with time-varying covariates for the survival endpoint and each latent class represents certain pattern of longitudinal and event-time responses. Larsen (2004) extended the Cox model to encompass a latent class variable (an indicator of the unobserved status of health or functioning) as predictor of time-to-event. However, the literature on the latent class model for survival data with a cure fraction is still sparse. Based on the subset of the data published in D'Amico et al. (2002), we showed in Section 5 that the proposed model with three latent cure rate markers fits the data best based on LPML, DIC and the reversible jump of Green (1995). This finding is consistent with the prostate cancer literature, as the three risk groups are routinely used in the prostate cancer clinical practice.

Although the proportional hazards (PH) assumption is assumed for the cumulative hazard function $H(y)$ for noncured subjects in (5), the resulting survival function is not PH due to the nature of the mixture model. To examine the PH assumption, we first considered the generalized odds-rate hazards (GORH) model discussed in Banerjee et al. (2007). We then compared various GORH models for $H(y)$ based on the LPML and DIC measures to see whether a PH model for $H(y)$ is appropriate. The results, which are available in Kim, Xi and Chen (2009), empirically confirm that the PH assumption for $H(y)$ may be appropriate for the prostate cancer data discussed in Section 1.

In Section 5, the covariates considered include only PSA, biopsy Gleason score, clinical tumor category and year of RP due to the limitation of the prostate cancer data we had. However, it will not add much additional computational difficulty to incorporate more covariates into the proposed model. Unlike D'Amico et al. (1998, 2002), the proposed model does not require any prespecified cutoff values of the covariates in classifying patients into different risk groups. The proposed method is potentially useful in clinical applications as it allows doctors to include as many important covariates as possible, some of which may be discovered later on due to medical advances, for obtaining a more accurate risk classification.

In the LCRM model, we assume that there are $G$ unknown latent $\theta_{g_i}$'s. Instead of the latent class model, we may assume a mixture of the Dirichlet Process (MDP) model discussed in Ibrahim, Chen and Sinha (2001) for the cure rate parameters. Specifically, we assign an unknown $\theta_i$ to each subject and then assume a Dirichlet Process prior for $\theta_i$. In Section 2.2, we assume a piecewise exponential model for the baseline hazard function $h_0(y)$. One possible extension to this is to assume a gamma process prior for $h_0(y)$, which leads to a semiparametric LCRM model. These two extensions of the LCRM model are currently under investigation.

## APPENDIX: PROOFS OF THEOREMS

PROOF OF THEOREM 1. After summing out $\mathbf{N}$ and $\mathbf{g}$, we have

$$\pi(\boldsymbol{\beta}, \boldsymbol{\theta}, \boldsymbol{\phi}, \boldsymbol{\lambda}|D_{obs}) \propto \pi^*(\boldsymbol{\beta}, \boldsymbol{\theta}, \boldsymbol{\phi}, \boldsymbol{\lambda}|D_{obs})$$
$$= L(\boldsymbol{\beta}, \boldsymbol{\theta}, \boldsymbol{\phi}, \boldsymbol{\lambda}|D_{obs})$$
$$\times \left[\prod_{j=1}^{J}\lambda_j^{-1}\right]\left[\prod_{k=1}^{G}\theta_k^{a_k-1}\exp(-b_k\theta_k)\right]\pi(\boldsymbol{\phi}),$$

where $L(\boldsymbol{\beta}, \boldsymbol{\theta}, \boldsymbol{\phi}, \boldsymbol{\lambda}|D_{obs})$ is given by (10). It suffices to show that

$$(A.1) \qquad \int \pi^*(\boldsymbol{\beta}, \boldsymbol{\theta}, \boldsymbol{\phi}, \boldsymbol{\lambda}|D_{obs})\, d\boldsymbol{\beta}\, d\boldsymbol{\theta}\, d\boldsymbol{\phi}\, d\boldsymbol{\lambda} < \infty.$$

It is easy to show that

$$L(\boldsymbol{\beta}, \boldsymbol{\theta}, \boldsymbol{\phi}, \boldsymbol{\lambda} | D_{obs})$$

$$\leq \prod_{\{i:\nu_i=1\}} \theta_G \exp\left\{ \mathbf{x}_i'\boldsymbol{\beta} + \sum_{j=1}^{J} \delta_{ij} \log \lambda_j - \exp(\mathbf{x}_i'\boldsymbol{\beta}) H_0^*(y_i) \right\}.$$

Using condition (iv), we can show

$$\int \theta_G^d \left[ \prod_{k=1}^{G} \theta_k^{a_k-1} \exp(-b_k\theta_k) \right] \pi(\boldsymbol{\phi}) \, d\boldsymbol{\theta} \, d\boldsymbol{\phi} < \infty$$

due to the constraints, $0 < \theta_1 < \theta_2 < \cdots < \theta_G$, and the condition, $a_k > 0$ and $b_k \geq 0$, for $k = 1, 2, \ldots, G-1$. Let

$$\pi^*(\boldsymbol{\beta}, \boldsymbol{\lambda} | D_{obs}) = \left[ \prod_{\{i:\nu_i=1\}} \exp\left\{ \mathbf{x}_i'\boldsymbol{\beta} + \sum_{j=1}^{J} \delta_{ij} \log \lambda_j - \exp(\mathbf{x}_i'\boldsymbol{\beta}) H_0^*(y_i) \right\} \right]$$

$$\times \left[ \prod_{j=1}^{J} \lambda_j^{-1} \right].$$

In order to establish (A.1), we only need to prove

(A.2) $$\int \pi^*(\boldsymbol{\beta}, \boldsymbol{\lambda} | D_{obs}) \, d\boldsymbol{\beta} \, d\boldsymbol{\lambda} < \infty.$$

Consider the transformation $u_j = \log(\lambda_j)$, and let $\mathbf{u} = (u_1, \ldots, u_J)'$. Then, $d\lambda_j = \lambda_j du_j$, $j = 1, 2, \ldots, J$, and

$$\pi^*(\boldsymbol{\beta}, \mathbf{u} | D_{obs}) = \pi^*(\boldsymbol{\beta}, \boldsymbol{\lambda} | D_{obs}) \left| \frac{\partial(\lambda_1, \lambda_2, \ldots, \lambda_J)}{\partial(u_1, u_2, \ldots, u_J)} \right|$$

$$= \prod_{j=1}^{J} \prod_{\{i:\nu_i=1\}} \left( \{\exp(u_j + \mathbf{x}_i'\boldsymbol{\beta})\}^{\delta_{ij}} \right.$$

$$\times \exp\left\{ -\delta_{ij} \exp(\mathbf{x}_i'\boldsymbol{\beta}) \right.$$

$$\times \left[ \exp(u_j)(y_i - s_{j-1}) \right.$$

$$\left. \left. \left. + \sum_{l=1}^{j-1} \exp(u_l)(s_l - s_{l-1}) \right] \right\} \right).$$

Letting $\delta_{ij_i} = 1$ and $\delta_{ij} = 0$ for $j \neq j_i$, we have

$$\pi^*(\boldsymbol{\beta}, \mathbf{u} | D_{obs}) \leq \pi^{**}(\boldsymbol{\beta}, \mathbf{u} | D_{obs})$$

$$= \prod_{\{i:\,\nu_i=1\}} \exp(u_{j_i} + \mathbf{x}_i'\boldsymbol{\beta}) \exp\{-(y_i - s_{j_{i-1}}) \exp(u_{j_i} + \mathbf{x}_i'\boldsymbol{\beta})\},$$

and it suffices to show that $\int \pi^{**}(\boldsymbol{\beta}, \mathbf{u}|D_{obs}) \, d\boldsymbol{\beta} \, d\mathbf{u} < \infty$. We rewrite $\pi^{**}(\boldsymbol{\beta}, \mathbf{u}|D_{obs})$ as

$$\pi^{**}(\boldsymbol{\beta}, \mathbf{u}|D_{obs})$$
$$= \prod_{j=1}^{J} \prod_{\{i:\,\nu_i=1, j_i=j\}} \exp(u_j + \mathbf{x}_i'\boldsymbol{\beta}) \exp\{-(y_i - s_{j-1}) \exp(u_j + \mathbf{x}_i'\boldsymbol{\beta})\}$$
$$= \prod_{j \neq j^*} \prod_{\{i:\,\nu_i=1, j_i=j\}} \exp(u_j + \mathbf{x}_i'\boldsymbol{\beta}) \exp\{-(y_i - s_{j-1}) \exp(u_j + \mathbf{x}_i'\boldsymbol{\beta})\}$$
$$\times \prod_{\{i:\,\nu_i=1, j_i=g^*\}} \exp(u_j + \mathbf{x}_i'\boldsymbol{\beta}) \exp\{-(y_i - s_{j-1}) \exp(u_j + \mathbf{x}_i'\boldsymbol{\beta})\}.$$

Since $d_j \geq 1$, there exists $s_{j-1} < y_{i_j} \leq s_j$ for $j \neq j^*$. Thus,

$$\prod_{j \neq j^*} \prod_{\{i:\,\nu_i=1, j_i=j\}} \exp(u_j + \mathbf{x}_i'\boldsymbol{\beta}) \exp\{-(y_i - s_{j-1}) \exp(u_j + \mathbf{x}_i'\boldsymbol{\beta})\}$$
$$\leq K_1 \prod_{j \neq j^*} \exp(u_j + \mathbf{x}_{i_j}'\boldsymbol{\beta}) \exp\{-(y_{i_j} - s_{j-1}) \exp(u_j + \mathbf{x}_{i_j}'\boldsymbol{\beta})\},$$

where $K_1 > 0$ is a constant, and

$$\int \prod_{j \neq j^*} \prod_{\{i:\,\nu_i=1, j_i=j\}} \exp(u_j + \mathbf{x}_i'\boldsymbol{\beta}) \exp\{-(y_i - s_{j-1}) \exp(u_j + \mathbf{x}_i'\boldsymbol{\beta})\}$$
$$\times \left( \prod_{j \neq j^*} du_j \right)$$
$$\leq K_2 \prod_{j \neq j^*} \int_{-\infty}^{\infty} \exp(u_j + \mathbf{x}_{i_j}'\boldsymbol{\beta}) \exp\{-(y_{i_j} - s_{j-1}) \exp(u_j + \mathbf{x}_{i_j}'\boldsymbol{\beta})\} \, du_j$$
$$= K_2 \prod_{j \neq j^*} (y_{i_j} - s_{j-1})^{-1},$$

where $K_2 > 0$ is a constant. For $j = j^*$, without loss of generality, we assume $y_{i_1^*}, \ldots, y_{i_{p+1}^*} \in (s_{j^*-1}, s_{j^*}]$, and $X_{j^*}^*$, which has the $l$th row $(1, \mathbf{x}_{i_l^*}')$, $\ell = 1, \ldots, p+1$, is of full rank. Therefore,

$$\prod_{\{i:\,\nu_i=1, j_i=j^*\}} \exp(u_{j^*} + \mathbf{x}_i'\boldsymbol{\beta}) \exp\{-(y_i - s_{j^*-1}) \exp(u_{j^*} + \mathbf{x}_i'\boldsymbol{\beta})\}$$
$$\leq K_3 \prod_{l=1}^{p+1} \exp\{u_j^* + \mathbf{x}_{i_l^*}'\boldsymbol{\beta}\} \exp\{-(y_{i_l^*} - s_{j^*-1}) \exp(u_j^* + \mathbf{x}_{i_l^*}'\boldsymbol{\beta})\},$$

where $K_3 > 0$ is a constant. Now consider the transformation $\mathbf{w} = (w_1, \ldots, w_{p+1})' \equiv X_{g^*}^* \binom{u_{j^*}}{\boldsymbol{\beta}}$, which is a one-to-one transformation. We have

$$\int_{R^{p+1}} \prod_{l=1}^{p+1} \exp(u_{j^*} + \mathbf{x}'_{i_{l^*}} \boldsymbol{\beta}) \exp\{-(y_{i_l^*} - s_{j^*-1}) \exp(u_{j^*} + \mathbf{x}'_{i_{l^*}} \boldsymbol{\beta})\} \, du_{j^*} \, d\boldsymbol{\beta}$$

$$\propto \int_{R^{p+1}} \prod_{l=1}^{p+1} \exp(w_l) \exp\{-(y_{i_l^*} - s_{j^*-1}) \exp(w_l)\} \, dw_l$$

$$= \prod_{l=1}^{p+1} (y_{i_l^*} - s_{j^*-1})^{-1}.$$

Therefore,

$$\int_{R^{p+J}} \pi^{**}(\boldsymbol{\beta}, \mathbf{u}|D_{obs}) \, d\boldsymbol{\beta} \, d\mathbf{u} \leq K \left( \prod_{j \neq g^*} (y_{i_j} - s_{j-1})^{-1} \right) \left( \prod_{l=1}^{p+1} (y_{i_l^*} - s_{j^*-1})^{-1} \right)$$

$$< \infty,$$

where $K > 0$ is a constant. This completes the proof. □

PROOF OF THEOREM 2. It is sufficient to show that $P(g_{new} = k|\boldsymbol{\beta}, \boldsymbol{\theta}, \boldsymbol{\phi}, \boldsymbol{\lambda}, t, \mathbf{x}_{new}, \mathbf{z}_{new}, G)$ for $k = 1$ $(k = G)$ increases (decreases) in $t$. We can rewrite the conditional predictive probability in (18) as

$$P(g_{new} = k|\boldsymbol{\beta}, \boldsymbol{\theta}, \boldsymbol{\phi}, \boldsymbol{\lambda}, t, \mathbf{x}_{new}, \mathbf{z}_{new}, G)$$

$$= \frac{\exp(\mathbf{z}'_{new} \boldsymbol{\phi}_k)}{\sum_{l=1}^G \exp(\mathbf{z}'_{new} \boldsymbol{\phi}_l) \exp(-(\theta_l - \theta_k)[1 - \exp\{-\exp(\mathbf{x}'_{new} \boldsymbol{\beta}) H_0^*(t)\}])}.$$

Since $H_0^*(t)$ is an increasing function of $t$, $\theta_l - \theta_1 > 0$ for $l > 1$, and $\theta_l - \theta_G < 0$ for $l < G$. Thus, $\hat{p}(k|t, \mathbf{x}_{new}, \mathbf{z}_{new}, G)$ increases in $t$ for $k = 1$ and decreases in $t$ for $k = G$. □

**Acknowledgments.** The authors wish to thank the Editor, the Associate Editor and three referees for helpful comments and suggestions which have improved the paper. The authors also wish to thank Dr. Anthony V. D'Amico of Brigham and Women's Hospital and Dana Farber Cancer Institute for his useful discussions and providing the prostate cancer data and the references on risk groups.

## SUPPLEMENTARY MATERIAL

**Checking the proportional hazards assumption and computational development** (DOI: 10.1214/08-AOAS238SUPP; .pdf). In online supplementary material we provide the empirical results for checking the proportional hazards assumption and the description of the Markov chain Monte Carlo (MCMC) sampling algorithm for a fixed $G$ and the detailed development of the reversible jump MCMC.

S. KIM
DIVISION OF EPIDEMIOLOGY
STATISTICS AND PREVENTION RESEARCH
NATIONAL INSTITUTE OF CHILD HEALTH
    AND HUMAN DEVELOPMENT, NIH
ROCKVILLE, MARYLAND 20852
USA
E-MAIL: kims2@mail.nih.gov

Y. XI
BIOGEN IDEC INC.
14 CAMBRIDGE CENTER
CAMBRIDGE, MASSACHUSETTS 02142
USA
E-MAIL: yingmei.xi@biogenidec.com

M.-H. Chen
Department of Statistics
University of Connecticut
215 Glenbrook Road, U-4120
Storrs, Connecticut 06269-4120
USA
E-mail: mhchen@stat.uconn.edu